\documentstyle[preprint,aps]{revtex}
\begin{document}
\preprint{KIAS-P99109}
%\preprint{KNU-H9911}
\def\a{\alpha}
\def\b{\beta}
\def\p{\partial}
\def\m{\mu}
\def\n{\nu}
\def\s{\sigma}
\def\half{\frac{1}{2}}
\def\hatt{{\hat t}}
\def\hatx{{\hat x}}
\def\hatth{{\hat \theta}}
\def\hatta{{\hat \tau}}
\def\hatrh{{\hat \rho}}
\def\hatva{{\hat \varphi}}
\def\p{\partial}
\def\nn{\nonumber}
\def\cb{{\cal B}}
\def\beq{\begin{eqnarray}}
\def\eeq{\end{eqnarray}}

\title{Noncommutative Dirac-Born-Infeld Action for D-brane}
\author{Taejin Lee
\thanks{E-mail: taejin@cc.kangwon.ac.kr}}
\address{{\it 
	Department of Physics, Kangwon National University, 
 	Chuncheon 200-701, Korea}}
\date{\today}
\maketitle
\begin{abstract}

We derive the noncommutative Dirac-Born-Infeld
action for the $D$-brane, which governs dynamics of $D$-brane with a
NS-NS $B$-field in the low energy regime. 
Depending on some details of the path integral prescriptions, 
both ordinary Dirac-Born-Infeld action and noncommutative one can be obtained
by evaluating the same Polyakov string path integral for the open string
ending on the $D$-brane. Thus, it establishes the equivalence
of the noncommutative Dirac-Born-Infeld action and the ordinary one.

\end{abstract}


\narrowtext

\section{Introduction}

The Dirichlet brane \cite{pol}, abbreviated to $D$-brane, is 
considered as one of the most important physical object 
to understand various aspects of
string theories. It has been the key ingredient to the subjects of 
dualities \cite{dual}, black hole physics \cite{black}, 
$AdS/CFT$ correspondence \cite{ads}, and 
Matrix $M$-model in modern string theory \cite{mtheo}. 
Before the advent of the $D$-brane, Fradkin and Tseytlin \cite{frad}
computed the effective action for an open string coupled to $U(1)$ 
gauge field and found that it is given by the Born-Infeld action at 
the tree level. Later Leigh \cite{leigh} studied the sigma model action 
for an open string in the $D$-brane background by requiring the conformal
invariance and found that the effective action for the $D$-brane 
should be the Dirac-Born-Infeld (DBI) action in the low energy regime. 
Then the DBI action has been often adopted to discuss 
the diverse subjects in string
theory in which the $D$ brane plays an essential role. 
The $Dp$-brane is the $(p+1)$ dimensional
hypersurfaces in space-time where the open strings can end and its 
dynamics is induced mostly by the open strings attached on it. 
The open string gives rise to the noncommutative geometry\cite{nc} for
the $D$-brane when a NS-NS $B$-field is present. The $D$-brane dynamics
is then described by Yang-Mills gauge fields on noncommutative space-time 
\cite{ncdb}. Most recently
Seiberg and Witten \cite{seib} proposed an explicit relationship
between the ordinary gauge fields and noncommutative gauge fields 
and in particular the equivalence of the ordinary Dirac-Born-Infeld 
action and the noncommutative one.

In the present paper we derive the noncommutative Dirac-Born-Infeld
action \cite{ncdbi} for the $D$-brane, which governs dynamics of $D$-brane 
with a NS-NS $B$-field. We show that both ordinary DBI action and 
noncommutative one can be obtained by evaluating the same Polyakov string 
path integral for the open string ending on the $D$-brane. The difference
in derivation of two DBI action only reside in some details of the 
path integral prescriptions. Thus, it establishes the equivalence
of the noncommutative Dirac-Born-Infeld action and the ordinary one.
The ordinary DBI action would be obtained \cite{frad} 
if we employ the Neumann function as the Green function on the disk 
and treat the terms involving the NS-NS $B$-field and the $U(1)$ 
gauge field as interaction. When $B$-field is constant, 
we do not need to treat the
term involving $B$-field as interaction. We may include this term in the
kinetic part of the action, quadratic in string variables, and define the
Green function with respect to it. In this case it is useful to employ
our previous canonical analysis \cite{tlee}: The end points of the string,
where the $U(1)$ gauge field is coupled to, obey noncommutativity and the
classical action becomes equivalent to that of open string in the
space-time with some effective metric, $G_E$. It suggests us that
we may get the noncommutative DBI action where the space-time
metric is replaced by the effective one $G_E$ and the ordinary
$U(1)$ field strength by its noncommutative counterpart.
We may include the term with $B$-field partly in the kinetic
part and partly in the interaction term. Then,
the derivation to be presented in this paper also suggests more
general form of equivalence between the ordinary gauge fields
and the noncommutative gauge fields, which is similar to one
discussed in ref.\cite{seib}.

\section{Open String on D-Brane and DBI action}

We begin with a brief review of the work of Fradkin and Tseytlin \cite{frad}
on DBI action. The bosonic part of the classical action for an open 
string ending on a $Dp$-brane with a $B$-field is given by
\begin{eqnarray} \label{action}
I &=& I_1 + I_2 \nn\\
&=& \frac{1}{4\pi\alpha^\prime}\int_M d^2 \xi 
\left[G_{\m\n} \sqrt{h} h^{\alpha\beta} \frac{\p X^\m}{\p \xi^\a}
\frac{\p X^\n}{\p \xi^\b} -2\pi i\a^\prime B_{ij} \epsilon^{\a\b}
\frac{\p X^i}{\p \xi^\a}\frac{\p X^j}{\p \xi^\b} \right]
\end{eqnarray}
where $\m = 0, 1, \dots, 9$ and $i=0, 1, \dots, p$ and 
$(\xi^0, \xi^1) = (\tau, \s)$.
Here $G_{ij}= g_{ij}=$constant, $H=dB=0$ and $h_{\a\b} = \delta_{\a\b}$.
Since the longitudinal string variables $X^\mu$, $\m=p+1, \dots, 9$
can be treated rather trivially, we will be concerned with the
transverse variables $X^i$ only afterwards.
At the tree level, the world surface of the open string is
a disk on the $D$-brane.
The interaction with $U(1)$ gauge field is introduced through
a Wilson loop defined on $\p M$, the boundary of the world surface, 
\beq
W[A] = P \exp\left(- i \int_{\p M} 
d{\hat\tau} A_i(X) {\dot X}^i \right)
\eeq
where ${\hat \tau}$ is the parameter along $\p M$
and $P$ denotes the path ordered product.
To be explicit, we choose ${\hat \tau}$ as
\beq
{\hat \tau} =
\left\{\begin{array}{r@{\quad:\quad}l}
\tau-1 &  {\hat \tau} \in [-1,0] \\
-\tau+1 & {\hat \tau} \in [0,1]  .
\end{array} \right. 
\eeq

The effective action for the $D$-brane is given as the Polyakov string
path integral on the disk
\beq
\Gamma = \frac{1}{g_s} N \int D[X] \exp\left(-I \right) W[A] \label{eff}
\eeq
where $g_s$ is the string coupling constant and $N$ is a normalization
constant. 
Using the Stokes theorem we may write the Wilson loop operator in the
string path integral as
\beq
W[A] = \exp \left(
- \frac{i}{2}\, \int_M \, d^2 \xi\, F_{ij}  \epsilon^{\a\b}
\frac{\p X^i}{\p \xi^\a}\frac{\p X^j}{\p \xi^\b} \right).
\eeq
For a slow varying $U(1)$ gauge field or a constant $F_{ij}$
\beq
W[A] = \exp \left(
- \frac{i}{2}\, \int_{\p M} d{\hat \tau} F_{ij}
X^i \frac{\p X^j}{\p {\hat \tau}} \right) = \exp(-I_3).
\eeq
For constant $B$-field we also write 
\beq
I_2 = -\frac{i}{2} \int_M B_{ij} \epsilon^{\a\b} 
\frac{\p X^i}{\p \xi^a} \frac{\p X^j}{\p \xi^\b} =
- \frac{i}{2}\, \int_{\p M} d{\hat \tau} B_{ij}
X^i \frac{\p X^j}{\p {\hat \tau}}.
\eeq
Therefore,
\beq
I_2+I_3 = - \frac{i}{2}\, \int_{\p M} d{\hat \tau} (B+F)_{ij}
X^i \frac{\p X^j}{\p {\hat \tau}}.
\eeq
Note that $(B+F)$ is invariant under the gauge transformation
where
\beq
A \rightarrow A+ \Lambda, \quad
B \rightarrow B -d\Lambda
\eeq
for any one-form $\Lambda$.

In order to evaluate the path integral it is convenient to diagonalize
the space-time metric, introducing
\beq
X^i_n = C^i{}_j Z^j_n, \quad (C^T g C)_{ij} = \delta_{ij}
\eeq
It follows that
\beq
\int D[X] &=& \int \prod_{n \ge 0, i} dX^i_n 
= \int \prod_{n \ge 0, i} dZ^i_n  \prod_{n \ge 0} \det C 
= \int D[Z] (\det g)^{\frac{1}{4}}.
\eeq
where we use $\zeta(0) = \sum_{n} 1 = -1/2$ and
$\det C = (\det g)^{-\half}$.
Here we note that according to the canonical analysis \cite{tlee}
the string normal modes, $X^i_n$ are subject to the following constraints
for the free open string
\beq
X^i_n = X^i_{-n}, \quad n= 1, 2, \dots.
\eeq
In evaluating the path integral we may treat $I_2$ and $I_3$ 
as interaction terms. It implies that the Green function on the 
disk is chosen as the Neumann function
\beq
- \frac{\p}{\p \xi^\a} h^{\a\b} \frac{\p}{\p \xi^\b} N_G = 
\delta(\xi-\xi^\prime).
\eeq
Thus, the path integral may be written as
\beq
\Gamma &=& \frac{1}{g_s} N (\det g)^{\frac{1}{4}} \int D[Z]
\exp \left[-I_1 - I_2 - I_3 \right] \\
&=& \frac{1}{g_s} N (\det g)^{\frac{1}{4}} \int d^{p+1} x \,
\int [dz] \nn\\ & & \quad
\exp \left[ -\half z G^{-1} z + i \pi \a^\prime
\left(C^T(B+F)C\right)_{ij} \int_{\p M} 
d{\hat \tau} \frac{\p z^i}{\p {\hat \tau}}
z^j \right] \nn
\eeq
where $z^i = Z^i \vert_{\p M}$ and
\beq
G({\hat \tau}_1,{\hat \tau}_2) = N_G(z({\hat \tau}_1), z({\hat \tau}_2)),\quad
G^{-1} G = \delta({\hat \tau}_1-{\hat \tau}_2).
\eeq
Employing the result of ref.\cite{frad}, we have
\beq
\Gamma &=& \frac{1}{g_s} N (\det g)^{\frac{1}{4}}
\int d^{p+1} x \, \sqrt{\det\left(
I + 2\pi \a^\prime C^T(B+F)C \right)} \\
&=& \frac{1}{g_s (2\pi)^p (\a^\prime)^{\frac{p+1}{2}}
(\det g)^{\frac{1}{4}}}\int d^{p+1} x \, \sqrt{\det\left(
g + 2\pi \a^\prime (B+F) \right)} \nn
\eeq
where we choose $N = 1/(2\pi)^p$.
Absorbing the factor $(\det g)^{\frac{1}{4}}$ into the string
coupling constant, we get the DBI Lagrangian
\beq
L_{DBI} = \frac{1}{g_s (2\pi)^p (\a^\prime)^{\frac{p+1}{2}}} 
\sqrt{\det\left(
g + 2\pi \a^\prime (B+F) \right)}.
\eeq

\section{Noncommutative Geometry}

When we derive the DBI action, we treat the term with the $B$-field
as interaction. However, since it is quadratic in string variables 
if $B$-field is constant, we may include it in the kinetic part of
the action. The Green function is defined with respect to $I_1+ I_2$
instead of $I_1$. Our previous canonical analysis \cite{tlee} 
shows that in this case the open string action is equivalent
to that of free open string in the space time with metric given by
$G_E$,
\beq
(G_E)_{ij} = \left(g - (2\pi\a^\prime)^2 Bg^{-1}B \right)_{ij}. \label{effg}
\eeq
The Hamiltonian and the string coordinate variable are
written in the phase space $(Y^i_n, K^i_n)$ by
\begin{mathletters}
\label{str:all}
\beq
H &=& (2\pi\a^\prime) \half p_i(G_E^{-1})^{ij}p_j
+(2\pi\a^\prime) \sum_{n=1}\left\{ \half K_{in} (G^{-1}_E)^{ij}
K_{jn} -\frac{1}{(2\pi\a^\prime)^2} \frac{n^2}{2} Y^i_n (G_E)_{ij} Y^j_n 
\right\}, \label{str:a}\\
X^i(\s) &=& x^i+i \theta^{ij}p_j \left(\s - \frac{\pi}{2}\right)+
\sqrt{2} \sum_{n=1} \left(Y^i_n \cos n\s + \frac{i}{n}
\theta^{ij} K_{jn} \sin n\s \right) \label{str:b}
\eeq
\end{mathletters}
where $Y^i_n$ and $K^i_n$ satisfy the usual commutation relation
\beq
\{Y^i_n, Y^j_m \} =0, \quad
\{Y^i_n, K_{jm} \} = \delta^i{}_j \delta_{nm}, \quad
\{K_{in}, K_{jm} \} = 0
\eeq
and
\begin{mathletters}
\label{theta:all}
\beq
\theta^{ij} &=& - (2\pi\a^\prime)^2
\left(\frac{1}{g+ 2\pi\a^\prime B} B\frac{1}{g- 2\pi\a^\prime B}
\right)^{ij} \label{theta:a} \\
(G_E^{-1})^{ij} &=& \left(\frac{1}{g+2\pi\a^\prime B}
g\frac{1}{g-2\pi\a^\prime B} \right)^{ij}. \label{theta:b}
\eeq
\end{mathletters}
In this representation, it is clear that the string variables
are noncommutative. In particular, the ends points of the open string
\begin{mathletters}
\label{end:all}
\beq
z^i &=& X^i(0) = x^i - \frac{\pi}{2}i \theta^{ij}p_j +
		\sqrt{2} \sum_{n=1} Y^i_n, \label{end:a}\\
{\bar z}^i &=& X^i(\pi) = x^i+  \frac{\pi}{2}i \theta^{ij}p_j +
		\sqrt{2} \sum_{n=1} (-1)^n Y^i_n \label{end:b},
\eeq
\end{mathletters}
satisfy
\beq
[z^i, z^j ] = i \pi\theta^{ij}, \quad
[{\bar z}^i, {\bar z}^j ] = -i \pi\theta^{ij} \label{commu}.
\eeq

The vertex operators carrying momenta $k$ and ${\bar k}$ are associated with 
$e^{ik^i z_i}$ and $e^{i{\bar k}^i {\bar z}_i}$. 
Hence, their operator algebra are given as
\beq
e^{ik\cdot z} e^{iq \cdot z} &=& e^{- i\frac{\pi}{2} k_i \theta^{ij} {q}_j
} e^{i(k+q)\cdot z}, \nn\\
e^{ik\cdot {\bar z}} e^{i q \cdot {\bar z}} &=& 
e^{i\frac{\pi}{2} k_i \theta^{ij} {q}_j} 
e^{i(k+q)\cdot {\bar z}}, \\
e^{ik\cdot z} e^{i q \cdot {\bar z}} &=& 
e^{i k\cdot z+i q \cdot {\bar z}}  \nn
\eeq
where we make use of the identity
\beq
e^A e^B = e^{\half [A,B]} e ^{A+B}, \quad {\rm if}\,\,\,
[[A,B],A] = [[A,B],B] = 0.
\eeq
The above noncommutative relations yield that the normal ordered
product of two operators are given as the Moyal bracket \cite{fair} 
as discussed in \cite{seib}. In general, a product of two functions 
of $z$ is written as
\beq
f(z) g(z) = \int \frac{dk}{2\pi} \int \frac{dq}{2\pi}
e^{- i\frac{\pi}{2} k_i \theta^{ij} {q}_j} e^{i(k+q)\cdot z}
{\tilde f}(k) {\tilde g}(q),
\eeq
where ${\tilde f}$ and ${\tilde g}$ are Fourier transformed functions of
$f$ and $g$ respectively.
It follows that normal ordered product of two operators satisfy
\beq
:f(z): :g(z): &=& :f(z) * g(z): 
\eeq
where
\beq
f(z) * g(z) &\equiv& \left. \exp\left(i \frac{\pi}{2} \theta^{ij}
\frac{\p}{\p \xi^i} 
\frac{\p}{\p \zeta^j}\right) f(z+\xi) g(z+\zeta)\right|_{\xi=\zeta=0}.
\eeq

The physical observables are often represented by Wilson loop operators.
Consider a Wilson loop operator of $U(1)$ gauge field on the $D$-brane,
given as follows
\beq
W_C[A] = P \exp \left[ \oint_C dX^i A_i (X) \right]
\eeq
where $P$ denotes the path ordered product.
Let us take that $C$ is the boundary of the world surface of the
open string on the $D$-brane, $\p M$ and is parameterize by
${\hat \tau}$:
\beq
X^i({\hat \tau}) \vert_{\p M} =
\left\{\begin{array}{r@{\quad:\quad}l}
z^i &  {\hat \tau} \in [-1,0] \\
{\bar z}^i & {\hat \tau} \in [0,1]  
\end{array} \right. 
\eeq
and $X^i({\hat \tau}= -1) =X^i({\hat \tau}= 1)$.

With the commutation relations Eq.(\ref{commu}), we write 
the expectation value of the Wilson loop operator as
\beq
\left<W_C[A] \right> &=& \int dzd{\bar z} \prod_{i, n} dX^i_n dP^i_n J(B)
\exp\Biggl[ \frac{i}{2\pi}\int d\tau \Biggl\{\left(\frac{d z^i}{d \tau} z^j 
-\frac{d {\bar z}^i}{d \tau} {\bar z}^j\right) 
(\theta^{-1})_{ij} \nn\\
& & \quad - H + \dots \Biggr\}\Biggr] \,\, P \exp \left[ \oint_{\p M} 
d{\hat \tau}\frac{dX^i}{d{\hat \tau}} A_i (X) \right] \\
&=& \int dzd{\bar z}\, \prod_{i, n} dX^i_n dP^i_n J(B)
\exp \left[\frac{i}{2\pi} \oint_{\p M} d{\hat \tau} 
\frac{dX^i}{\p {\hat\tau}} X^j (\theta^{-1})_{ij} - 
i\int d\tau H + \dots \right] \nn\\
& & \quad \exp \left[ \oint_{\p M} 
d{\hat \tau}\frac{dX^i}{d{\hat \tau}} A_i (X) \right] \nn
\eeq
where $J(B)$ is a trivial Jacobian and 
`$\dots$' denotes the kinetic terms for nonzero modes and 
constraint terms.
Note the difference between the $\tau$ ordered product and the
path ordered product. If ${\hat \tau}$-ordering is employed, on ${\p M}$
\beq
e^{iP\cdot X} e^{iQ \cdot X} 
&=& e^{- i\frac{\pi}{2} P_i \theta^{ij} {Q}_j} e^{i(P+Q)\cdot X} =
e^{iP\cdot X} * e^{iQ \cdot X}. \label{tau}
\eeq
We may expand the Wilson loop operator as
\beq
W_C[A] = I + \int_{\p M} dX \cdot A + \half
{\int_{\p M} \int_{\p M}}_{X_2 > X_1} \,\,\,dX_2 \cdot A(X_2)
dX_1 \cdot A(X_1) + \dots. 
\eeq
Expanding $A_i[X({\hat \tau})]$ also and using Eq.(\ref{tau}) we get
\begin{mathletters}
\label{wil:all}
\beq
\left< W_C[A] \right>&=& \left<I + 
\int_{M} d\tau d\s \left(\frac{\p Y^i}{\p \tau}
\frac{\p Y^j}{\p \s} - \frac{\p Y^i}{\p \s}
\frac{\p Y^j}{\p \tau}\right) \hat F_{ij} \right> + \dots, \label{wil:a}\\
{\hat F}_{ij} &=& \p_i {\hat A}_j - \p_j {\hat A}_i - 
{\hat A}_i * {\hat A}_j + {\hat A}_i * {\hat A}_j, \label{wil:b} \\
{\hat A}_i &=& A_i - \frac{\pi}{4} \theta^{kl} \{ A_k, \p_l A_i+ F_{li} \}_S
+ {\cal O}(\theta^2) \label{wil:c}
\eeq
\end{mathletters}
where
\beq
Y^i(\s) = x^i + \sqrt{2} \sum_{n=1} Y^i_n e^{in\s}. 
\eeq
Note that
\beq
d X \cdot A = d\left(x^i+ \sqrt{2}\sum_{n=1} Y^i_n\right) A_i
+\frac{\pi}{2}i dp_i \theta^{ij} A_j.
\eeq
When we evaluate the expectation value of the Wilson loop operator,
we may take the Wick contraction between $p$ and $A$. This procedure
turns $A$ into its noncommutative counterpart ${\hat A}$. 
A similar argument can be found in ref.\cite{oku}.
If we apply the (non-)Abelian Stokes theorem to the case of 
noncommutative algebra, we may find
\beq
\left<W_C[A] \right>
= \left<\exp*\left[\int_M d\tau d\s \left(\frac{\p Y^i}{\p \tau}
\frac{\p Y^j}{\p \s} - \frac{\p Y^i}{\p \s}
\frac{\p Y^j}{\p \tau}\right) \hat F_{ij}\right]\right>
\eeq
where
\beq
\exp* (A) = \sum_n \frac{1}{n!} (A*A* \dots* A).
\eeq
The noncommutative Stokes theorem needs a more rigorous proof.

\section{Noncommutative DBI action}

Being equipped with the canonical analysis \cite{tlee} and the discussion 
on the noncommutative geometry given in the previous section, 
we evaluate the Polyakov string path integral. With the prescription given
in the previous section the Polyakov string path integral 
representing the effective action is read as
\beq
\Gamma &=& \frac{1}{g_s} N \int D[X] 
\exp\left(-I_1 - I_2 \right) W[A] \label{effact} \\
&=& \frac{1}{g_s} N \int D[Y, K] \exp\left[\int d\tau \left(p_i \dot{
x}^i + \sum_n K_{in}\dot{Y}^i_n - H \right) \right] W[A] \nn
\eeq
where $W[A] = P \exp\left( -i\int_{\p M} d{\hat \tau} A_i(X) \dot{X}^i
\right)$.
As discussed if we include the $B$-field term in the kinetic part of the
action, the Wilson loop operator in the Polyakov path integral 
may be rewritten as
\beq
\left< W[A] \right> = \left< \exp\left[-\frac{i}{2} 
\int _M d^2\xi \, {\hat F}_{ij} 
\epsilon^{\a\b} \frac{\p Y^i}{\p \xi^\a}
\frac{\p Y^j}{\p \xi^\b} \right] \right>. \label{wilson}
\eeq
Following Seiberg and Witten \cite{seib}, we substitute
ordinary products for the $*$ products between ${\hat F}$ in 
Eq.(\ref{wilson}), since it makes difference only in terms 
with derivatives of ${\hat F}$. For a slow varying $U(1)$ gauge 
field, we may also write the Wilson loop operator as
\beq
\left< W[A] \right> = \left< \exp \left(
- \frac{i}{2}\, \int_{\p M} d{\hat \tau} {\hat F}_{ij}
Y^i \frac{\p Y^j}{\p {\hat \tau}} \right) \right>
= \left< \exp \left(-{\hat I}_3 \right) \right>.
\eeq

Integrating out the momentum variables in Eq.(\ref{effact}),
\begin{mathletters}
\label{non:all}
\beq
\Gamma &=& \frac{1}{g_s} N \int D[Y]
\exp \left(-I_E-\frac{i}{2} \int _M d^2\xi \, {\hat F}_{ij} 
\epsilon^{\a\b} \frac{\p Y^i}{\p \xi^\a}
\frac{\p Y^j}{\p \xi^\b} \right), \label{non:a}\\
I_E &=& \frac{1}{4\pi\alpha^\prime}\int_M d^2 \xi 
(G_E)_{ij} \sqrt{h} h^{\alpha\beta} \frac{\p Y^i}{\p \xi^\a}
\frac{\p Y^j}{\p \xi^\b}, \label{non:b}
\eeq
\end{mathletters}
we find that the string path integral Eq.(\ref{non:a}) coincides with
Eq.(\ref{eff}), if the effective metric $G_{ij}$ substitutes
for the space-time metric $g_{ij}$ and the noncommutative field
strength ${\hat F}$ for $(B+F)$. Then, the same procedure which
leads to the ordinary DBI action, yields
\beq
\Gamma &=& \frac{1}{g_s (2\pi)^p (\a^\prime)^{\frac{p+1}{2}}
(\det G_E)^{\frac{1}{4}}} \int d^{p+1} x \, \sqrt{\det\left(
G_E + 2\pi \a^\prime {\hat F} \right)}
\eeq
Absorbing the factor $(\det g)^{-\frac{1}{4}}$ into the
string coupling constant $g_s$ as before, we arrive at
the noncommutative DBI Lagrangian 
\begin{mathletters}
\label{ncdbi:all}
\beq
{\hat L}_{DBI} &=& \frac{1}{G_s (2\pi)^p (\a^\prime)^{\frac{p+1}{2}}} 
\sqrt{\det\left(
G_E+ 2\pi \a^\prime {\hat F} \right)} \label{ncdbi:a}, \\
G_s &=&  g_s \left(\frac{\det G_E}{\det g}\right)^{\frac{1}{4}}
= g_s\left(\frac{\det(g+2\pi\a^\prime B)}{\det g}\right)^{\frac{1}{2}}.
\label{ncdbi:b} 
\eeq
\end{mathletters}

\section{Concluding Remarks}

In ref.\cite{seib} Seiberg and Witten discussed the equivalence
between the noncommutative gauge theory and the ordinary one
and the change of variables between them in an explicit from. 
The proposed equivalence was checked by comparing the noncommutative 
DBI with the ordinary one, both of which are supposed to describe
the same $D$-brane with a NS-NS $B$-field. In the present paper
we derive the noncommutative DBI action, evaluating the Polyakov
string path integral on a disk, which depicts the world surface
of the open string ending on the $D$-brane. We get both noncommutative
DBI and ordinary DBI from the same Polyakov string path integral.
Thus, it is established that the noncommutative DBI action is
equivalent to the ordinary one.
Some details of the prescriptions for the path integral make the
difference. If $B$-field is constant, the term involving the
$B$-field can be treated as a part of interaction or as a part of
kinetic term, since it is quadratic in string variables.
In the former case we get the ordinary DBI action and in the latter
case the noncommutative one. In ref.\cite{seib} two descriptions,
one by the ordinary gauge theory and the other by the noncommutative
one are shown to differ by the choice of regularization for 
the world-sheet theory; the Pauli-Villars regularization yields 
the ordinary commutative gauge symmetry while the point-splitting 
regularization yields the noncommutative one. The analysis of the 
string path integral in the present paper may be compared with 
theirs. 

Since whether the term with $B$-field is put in the
kinetic part or in the interaction part is optional, we may get
a more general form of the noncommutative DBI action. We may split the
term with the $B$-field into two and put one in the kinetic
part and the rest in the interaction part. Then the string path
integral will leads us to a more general form of the noncommutative
DBI action. Thus, our description of the noncommutative DBI 
provides a useful tool to examine the interesting proposal made
in \cite{seib} in some details. It is interesting to explore further 
its consequence \cite{tj}. The open strings attached to the 
multi-$D$-branes or to two different types of $D$-branes can be
treated in similar ways. It is certainly interesting to understand
the noncommutative non-Abelian DBI action in the framework presented
here.

After having completed the work I found that equivalence between the ordinary
DBI and the noncommutative DBI has been discussed also in refs. \cite{others}.

\section*{Acknowledgement}
This work was supported in part by KOSEF. 
Part of the work was done during the author's visit to 
PIMS (Canada) and KIAS (Korea).

\end{document}